**2D IR spectroscopy of catalytic monolayers**

**Proposal**

**by**

**Seyedeh Tahereh Alavi**

**Department of Chemistry**

**University of Missouri**

**Main objective:**

The main goal of this study is to compare the catalytic reactivity of a metal carbonyl complex in different environments (heterogeneous and homogeneous) by monitoring the spectral diffusion rate and vibrational lifetime. Since solvent addition to the surface of an immobilized catalyst or submerging the catalyst in bulk solution, changes the microscopic characteristics of the system, we need to get its dynamical information to investigate the catalytic reactivity. To achieve such information, we aim to use Two-Dimensional Infrared Spectroscopy (2D IR) which is a powerful multidimensional vibrational spectroscopy method that have been developed to study dynamical processes in condensed phase systems. For this study, we picked acetonitrile complex $MoI_2(CO)_3(NCCH_3)_2$ which is used to catalyze the polymerization reaction of functional acetylene in room temperature.

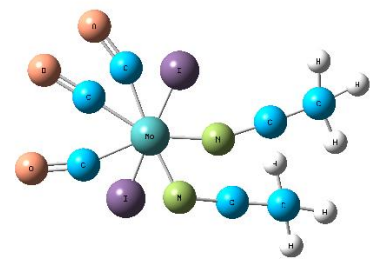

Figure 1: molecular structure of halogenated acetonitrile transition metal complex

**Introduction:**

Polyacetylene is an organic semiconductor that shows increased conductivity when being doped with halogens. Although organic conductors seem to be of high interest in microelectronics, polyacetylene is not commercially used because of instability in air which makes it difficult to be processed [1]. When hydrogen atoms in acetylene monomers are substituted by other functional groups, the resulted polymer gets more rigid and shows interesting properties such as photoconductivity, optical nonlinearity and liquid crystallinity[2]. Polyacetylene is commonly produced using Ziegler-Natta catalyst. This catalytic reaction that are based on metal insertion, are limited to non-substituted acetylene because functional group insertion can deactivate the catalyst. Furthermore, in the presence of acetylene monomers, polyacetylene formation is dominated because of relative electron deficiency of functionalized acetylene in the triple bond [3]. To circumvent this problem, inorganic and organometallic compounds such as Molybdenum and Tungsten halides have been used in substituted acetylene polymerization[4]. These catalysts have shown good reactivity but their instability in the presence of air and water makes them extremely expensive in terms of shipping and storage. Metal carbonyl complexes such as Molybdenum and Tungsten hexacarbonyl ($Mo(CO)_6$) which are not sensitive to air and water are good replacement for such catalysts, but they need to be activated by being irradiated with UV light [5]. Since in industry it is preferred not to use photochemical processes, metal carbonyl complexes with $Mo(CO)_xL_y$ structure are being used in which the organic ligand has a role of catalyst activation. But these complexes only catalyze polymerization reactions in halogenated solvents such as chloroform and does not have moisture stability. So, by reacting them with halogens such as iodine, the non-stability problem can be solved. Tang and coworkers used halogenated acetonitrile complexes ($MoI_2(CO)_3(NCCH_3)_2$) to polymerize phenyl acetylene in room temperature. Such complexes are able to lead the polymerization reaction in both halogenated and non-halogenated solvents with relatively high efficiency [2]. Because of difficulty in separation of products from such homogeneous catalysts, in some cases especially in pharmaceutical industries where the purification is essential, it is preferred to immobilize the catalyst on an organic or inorganic support to lower the expenses of separation and product purification. Immobilized catalysts show decreased activity because of lower interaction with the reactants [6]. Adding solvents to the surface can affect the molecular dynamic of the system and thus changes the reactivity. In this study we aim to investigate the effect of adding solvent to the surface of heterogeneous catalysts. We can compare the catalytic reactivity of heterogeneous catalyst of interest in wet and dry interfaces with that of the catalyst in bulk solution by studying the dynamic of the system.

In the dynamical study of liquid phase systems, we need to use a spectroscopy method capable of resolving all fast dynamical processes happening inside the system. Conventional one-dimensional spectroscopy methods have the required time resolution to study such fast processes. But, such spectra are usually very congested and difficult to be analyzed. Even if we can resolve all the peaks on a one- dimensional spectrum, it does not provide us with enough structural information needed to determine the dynamic of the system [7]. So, we need to use multidimensional spectroscopy methods in which the response of a system is studied as a function of multiple axes. Two-Dimensional Nuclear Magnetic Resonance (2D NMR) which is a commonly used multidimensional spectroscopy method, has been used for more than four decades to study the dynamic of condensed phase systems. Using this method, one can separate the overlapped peaks which is the major problem on one dimensional spectra and obtain the detail structural information, such as connection or separation of nuclei by analyzing the cross peaks [8]. However, 2D NMR has a low time resolution. It can only observe processes that happen in the range of milliseconds. To address this problem, 2D IR spectroscopy method was developed by Isao Noda [9]. In this spectroscopic method three short IR pulses interact with a multi-level vibrational system and consequently a third order nonlinear signal is emitted. The timescale of the experiment is in the range of sub picoseconds which means it is fast enough to study any dynamical processes occurring in the system [10]. Such experiments are characterized by three time periods. During the first period (the evolution time $t_e$) a coherent superposition of states (wave packet) is produced. After second pulse (waiting time $t_w$) population starts and vibrational oscillators evolve to new frequencies. In the third time period (detection time $t_d$) the system emits a signal through stimulated emission. The signal is plotted as a function of frequencies associated with first and third pulses in a 2D IR correlation spectrum [11]. Each spectrum typically contains some diagonal and cross peaks depending on the system under study. Diagonal peaks arise from fundamental and overtone vibrational transitions, while cross peaks could be produced as results of coupling between vibrational oscillators, chemical exchanges or non-equilibrium chemical reactions [12].

Khalil *et al*. (2003) used polarization selective 2D IR spectroscopy to study the dynamic of RDC in hexane and chloroform in terms of two strongly coupled symmetric and asymmetric carbonyl group stretches [7]. They (2004) also characterized vibrational coherence transfer, population relaxation and dephasing inside the same system by investigating the relative peak amplitude and appearance of relaxation-induced peaks on 2D IR spectra [13]. Peng and Tokmakoff (2012) used 2D IR spectroscopy to separate lactim and lactam tautomers of aromatic heterocycles in water solutions [14]. 2D IR spectroscopy is also used in many biological studies where the conformational information and fast structural changes are determined. Krummel *et al*. (2003) used this method to study the coupling between cytidine and guanosine bases in DNA [15]. Another important application of 2D IR spectroscopy method is measuring the emitted signal from a monolayer of molecules at gas-solid and liquid-solid interfaces which is useful in the study of biological molecules and heterogeneous catalysts. Rosenfeld *et al*. (2011) used this method to investigate the interfacial dynamic of a photocatalyst in three different environments [16].

Here we intend to use this vibrational spectroscopy method to evaluate the catalytic reactivity of halogenated acetonitrile complexes in bare and wet interfacial monolayers as well as in bulk solution.

**Theory:**

Interaction of a radiation field with a system creates a non-equilibrium charge distribution called macroscopic polarization. In time-resolved experiments, we measure the macroscopic polarization by detecting the emitted field to extract the response of the system (the time evolution of wave packet) [12].

When a laser pulse interacts with the system, the density matrix evolves during the time under the influence of unperturbed and perturbed parts of the Hamiltonian:

$$\hat{H} = \hat{H}_0 - \mu E(t)$$

Here $\mu$ is the dipole transition moment and $E(t)$ is the electric field of the pulse. First the density matrix is multiplied by dipole transition moment from both sides (ket and bra sides) to get $\rho^1$ and then by taking the trace $<\mu(t_1)\rho^1>$ (time evolution of wave packet under the unperturbed part) the first order response function is obtained:

$$\rho^1 = i(\mu(0)\rho(-\infty) - \rho(-\infty)\mu(0))$$

$$R^{(1)}(t_1) = i\langle \mu(t_1)\,\mu(0)\rho(-\infty) \rangle - i<\rho(-\infty)\mu(0)\mu(t_1)\rangle$$

So, the macroscopic polarization is calculated by convoluting the first order response function with the electric field of the pulse (linear response theory):

$$P(t) = \int_0^\infty dt_1 E(t-t_1)R^{(1)}(t_1)$$

We can extend it to a third order nonlinear response where three radiation fields interact with the system (which is the case in 2D IR spectroscopy method) [12]:

$$R^{(3)}(t_3,t_2,t_1) =$$
$$i\langle \mu_3\,\mu_1\rho(-\infty)\mu_0\mu_2 \rangle - i\langle \mu_2\,\mu_0\rho(-\infty)\mu_1\mu_3 \rangle \Rightarrow R_1 + R_1^*$$
$$i\langle \mu_3\mu_2\rho(-\infty)\mu_0\,\mu_1 \rangle - i\langle \mu_1\,\mu_0\rho(-\infty)\mu_2\mu_3 \rangle \Rightarrow R_2 + R_2^*$$
$$i\langle \mu_3\,\mu_0\rho(-\infty)\mu_1\mu_2 \rangle - i\langle \mu_2\,\mu_1\rho(-\infty)\mu_0\mu_3 \rangle \Rightarrow R_4 + R_4^*$$
$$i\langle \mu_3\,\mu_2\mu_1\mu_0\rho(-\infty) \rangle - i\langle \rho(-\infty)\,\mu_0\mu_1\mu_2\mu_3 \rangle \Rightarrow R_5 + R_5^*$$

Here $\mu_i$ is the transition dipole moment after ith radiation-matter interaction, $\rho(-\infty)$ is the density matrix, $t_i$ is the ith time period and $R_i$ is the response. The combination of each response and its complex conjugate ($R_i$ and $R_i^*$), give a Liouville's pathway. The evolution of density matrix under pathways 1 is shown below:

$$\rho(t) = \begin{pmatrix} 1 & 0 \\ 0 & 0 \end{pmatrix} \xrightarrow{pulse} \begin{pmatrix} 0 & +i \\ -i & 0 \end{pmatrix} \xrightarrow{t_1} \begin{pmatrix} 0 & +ie^{+iw_{01}t_1} \\ -ie^{-iw_{01}t_1} & 0 \end{pmatrix} \xrightarrow{pulse} \begin{pmatrix} 0 & 0 \\ 0 & -\sin(w_{01}t_1) \end{pmatrix}$$

$$\xrightarrow{pulse} \begin{pmatrix} 0 & -ie^{+iw_{01}t_1} \\ +ie^{-iw_{01}t_1} & 0 \end{pmatrix} \xrightarrow{t_3} \begin{pmatrix} 0 & -ie^{+iw_{01}(t_3-t_1)} \\ +ie^{-iw_{01}(t_3-t_1)} & 0 \end{pmatrix} \xrightarrow{emit} \sin(w_{01}(t_3 - t_1))$$

Here $w_{01}$ is the fundamental transition frequency. Under this pathway the coherence that is started in the system after first laser pulse, is changed to its complex conjugate after second and third input pulses, which results in reappearance of macroscopic polarization in the systems with the inhomogeneous broadening. Such pathways are called photon echo or rephasing pulse sequence [12].

Unlike a rephasing pulse sequence, in non-rephasing pathways the coherence is not altered after second and third pulses. Pathway four is representative for such pathways:

$$\rho(t) = \begin{pmatrix} 1 & 0 \\ 0 & 0 \end{pmatrix} \xrightarrow{pulse} \begin{pmatrix} 0 & -i \\ +i & 0 \end{pmatrix} \xrightarrow{t_1} \begin{pmatrix} 0 & -ie^{+iw_{01}t_1} \\ +ie^{-iw_{01}t_1} & 0 \end{pmatrix} \xrightarrow{pulse} \begin{pmatrix} 0 & 0 \\ 0 & \sin(w_{01}t_1) \end{pmatrix}$$

$$\xrightarrow{pulse} \begin{pmatrix} 0 & -ie^{+iw_{01}t_1} \\ +ie^{-iw_{01}t_1} & 0 \end{pmatrix} \xrightarrow{t_3} \begin{pmatrix} 0 & -ie^{+iw_{01}(t_3+t_1)} \\ +ie^{-iw_{01}(t_3+t_1)} & 0 \end{pmatrix} \xrightarrow{emit} \sin(w_{01}(t_3 + t_1))$$

It is seen that in this case there is a continuous decrease in the macroscopic polarization. The key difference between rephasing and non-rephasing pathways is oscillation of Bloch vectors with conjugate frequencies during the first time periods. An absorptive 2D IR correlation spectrum is produced by adding equally weighted rephasing and non-rephasing spectra to eliminate the dispersive characteristics of the signal [7]. These two pathways are discriminated by phase matching conditions.

**Experimental setup:**

To immobilize the halogenated acetonitrile complex, it is bonded to a silica surface (grown on a CaF$_2$ window which is transparent to mid-IR beams) by tethering to a monolayer of hydrocarbon chain (such as undecane) through a stable linker [17] (such as triazole).

To study the dynamic of the catalyst, we need to get 2D IR correlation spectrum of the stretching mode in which all three carbonyl groups in the complex are engaged. Carbonyl stretching modes have strong absorption in the range of 1600 to 2100 cm$^{-1}$ depending on the groups attached to the carbon [18]. Since changes in the charge density on carbonyl groups in the metal complex alters their vibrational motions, we can relate the frequency shifts and peak narrowing of such modes to the environmental changes [16]. So, by analyzing the peaks line shape and spectral diffusion observed on 2D IR spectrum, we will be able to get information about the dynamic of each environment.

To do the experiment, first we need to generate short mid-IR pulses centering at around 2000 cm$^{-1}$ (5 $\mu m$) to be able to see the stretching mode of interest. We can produce such IR pulses with sub 100 fs temporal bandwidth, by passing the output of a Ti:Sapphire laser amplified by chirp pulse amplification method, operating at 1 kHz through an optical parametric amplifier (OPA). An OPA consists of three stages. In the first stage, the pump beam (second harmonic or fundamental output of Ti:Sapphire laser) is used to produce a seed beam. In ultrafast OPA two techniques are used for generating this beam: white light continuum or parametric super fluorescence. We will use a white light continuum method, in which the ultrafast beam is focused inside a transparent material such as sapphire plate and spectral broadening occurs [19]. The produced white light is temporally chirped, so, we can choose the wavelength of seed beam that we need by tuning the delay time. After producing the seed beam, it is combined with the pump beam in nonlinear crystals to generate the idler beam. To get maximum intensity of the idler, three beams must meet the phase matching condition throughout the propagation. The phase matching direction for these beams is written below [19]:

$$hk_p = hk_s + hk_i$$

Finally, the idler beam is passed through second harmonic generator and difference frequency mixing units to get the frequency we need for the experiment (Figure 2).

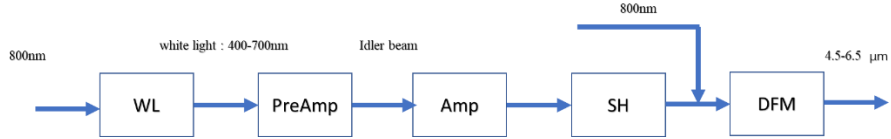

Figure 2: output of a Ti:Sapphire laser enters an OPA and after passing through pre-amplifier and amplifier, it goes through second harmonic generator (SHG) and difference frequency mixing (DFM) units to get the desired wavelength .

Like most 2D IR spectroscopy experiments we use a BoxCAR geometry (figure 3). In this experimental setup the produced IR beam is propagated through a five-beam interferometer [7]. The IR beam is split into five beams (three input pulses, a tracer and a local oscillator) through beam splitters.

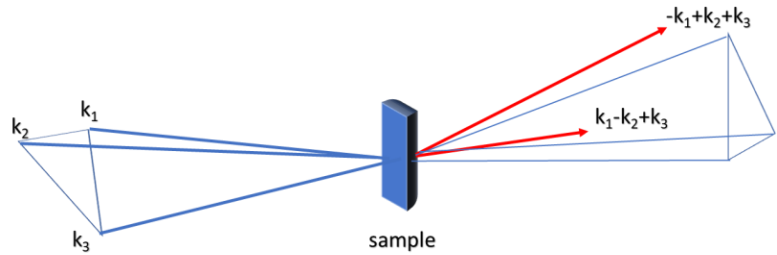

Figure 3: beams interact with the system in BoxCAR geometry.

Three input pulses are focused within the sample using a parabolic mirror in a BoxCAR geometry as shown in figure 4. The spatial overlap of three beams are achieved by passing them through a pinhole (the 50 μm hole is made on a shim metal plate using laser ablation method) at the focus of the parabolic mirror before the experiment. The relative timing between the beams can be controlled using pair of mirrors mounted on linear translational stages. Tracer beam that is only used for alignment and follows the expected pathway of the signal, is blocked during the experiment. The nonlinear signal emitted in the experiment is in the time domain and needs to be Fourier transformed. To do Fourier transformation, phase information is needed. To get these information it is temporally and spatially overlapped with a fixed local oscillator (LO) before being dispersed inside the monochromator and the interference of two beams is analyzed. The detection frequency is obtained from the monochromator which gives the vertical axis of the spectrum ($w_3$) and then the evolution time is scanned at each $w_3$ which makes the signal phase change with respect to LO and a temporal interferogram is obtained at each detection frequency. Finally, by numerical Fourier transformation of each temporal interferogram, the horizontal axis is gained [20].

The IR array detector is connected to a high-speed digitizer and data acquisition software. The data is recorded in the same repetition rate as the laser. The experiment uses automated control of the translational stages and data acquisition to get both rephasing and non-rephasing signals. This program can be written in LabView. According to rotating wave approximation (RWA) The rephasing signal is emitted in the $S = -k_1 + k_2 + k_3$ phase matching direction and non-rephasing signal is emitted in the $S = k_1 - k_2 + k_3$ direction [12]. To record both signals the time sequences of three pulses are changed (1-2-3 for the rephasing and 2-1-3 for non-rephasing signal).

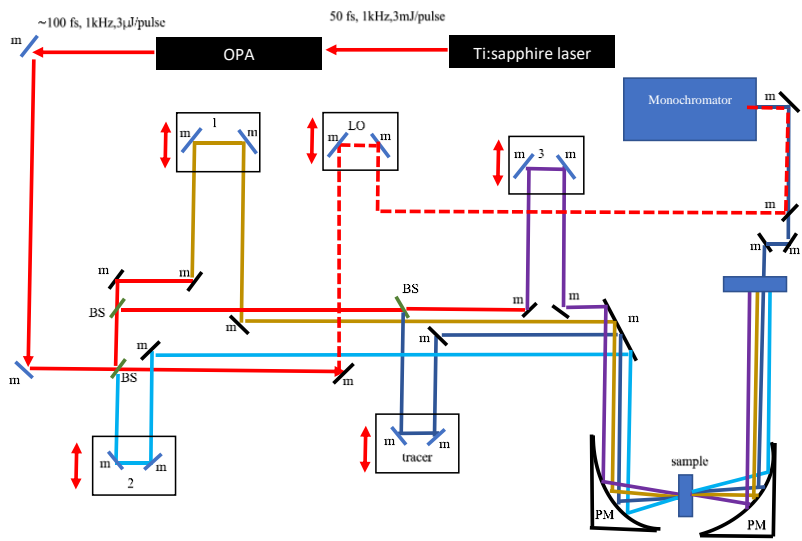

Figure 4: Experimental setup with BoxCAR geometry. M: mirror, PM: parabolic mirror, BS: beam splitter, LO: local oscillator

In a liquid phase system, each molecule experiences a slightly different environment compared to other molecules. So, even in pure liquid phase systems, instead of one single frequency for a specific vibrational mode, we expect to see a distribution of frequencies (inhomogeneous broadening). If we probe such systems in timescales much faster than the time needed for the environment to change, the peak on 2D IR spectrum would be diagonally elongated. We can measure the inhomogeneous broadening from diagonal linewidth of the peak and homogeneous broadening from antidiagonal linewidth (Figure 5) [12].

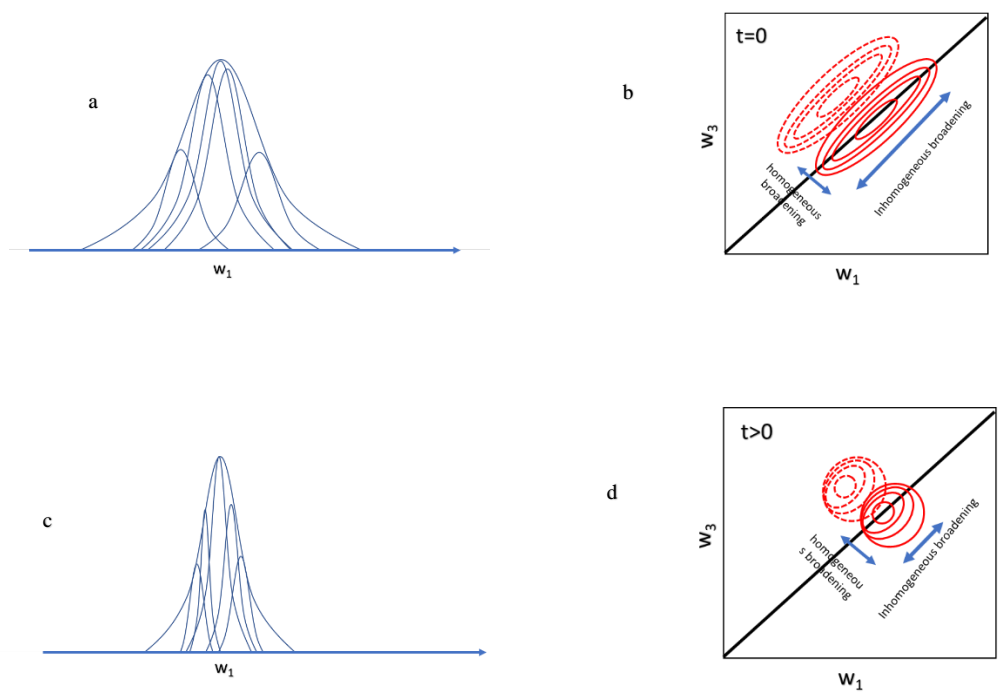

Figure 5: a:inhomogoneousely broadened vibrational transitions b: its resulting 2D IR correlation spectrum at $t_w$=0. The peak is diagonally elongated. c:inhomogeneous broadening decrease at $t_w$>0 d: its resulting 2D IR spectrum. The peak has a more symmetric shape.

This change of frequency distribution as a function of waiting time is called spectral diffusion that can be characterized by frequency-frequency correlation function (FFCF). FFCF which is a bridge connecting experimental observables to microscopic molecular dynamic, is defined as the probability that a vibrational oscillator has the same frequency as the initial one after some time t. FFCF can be obtained by calculating third order response function based on time-dependent perturbation theory repeatedly until it converges to the experimental observable. Park *et al*. (2007) described an easier way of extracting FFCF from 2D IR spectrum Instead of this time-consuming theoretical calculation. In this method a series of slices at each detection frequency (parallel to the horizontal axis) is cut through the 2D IR spectrum. These slices are projected on the horizontal axis and by fitting each one of them to Gaussian distribution, the peak positions are found. Then a linear fitting of slices' peak positions is performed to yield center line slope (CLS) which varies between 0 and 1. CLS is a good representative of the spectral diffusion rate that gives the inhomogeneous waiting time-dependent component of FFCF. As the waiting time increases, CLS increases toward 1 (center line gets vertical) [21].

At each set of experiment a 2D IR spectrum is obtained by adding the rephasing and non-rephasing spectra which are plotted by scanning the evolution time at each detection frequency. In the beginning, the spectrum at $t_w = 0$ is obtained and then by increasing the waiting time in 2 picoseconds interval more spectra are recorded. We repeat these experiments for the catalyst in bulk solution (hexane), immobilized on silica surface with solid-air and solid-liquid (hexane is added to the interfacial monolayer) interfaces.

So, to evaluate the dynamic of each environment, we can plot CLS as a function of waiting time for each of these environments. By fitting CLS to an exponential function we get a decay time constant. It is expected to have shorter decay time constant for the bulk solution compared to the heterogeneous catalyst which is indicative of faster structural evolution in the solution.

We can also get vibrational lifetime of the interested mode at each environment using a combination of 2D IR and heterodyne-detected transient grating spectroscopy (HDTG) [16]. In HDTG the constructive and destructive interferences of excitation pulses generate an excited state population that varies spatially and acts as a diffraction grating. The third beam passes through this diffraction pattern after some delay. Time-dependent diffraction of third beam is plotted to give the vibrational lifetime [22]. Here, we expect to see lower vibrational lifetime for the bulk solution compared to two other environments because of increasing the phonon density of states. Also, in the comparison of two bare and wet monolayers, the wet monolayer probably shows lower vibrational lifetime as a result of the same reason. The vibrational frequency also is expected to change by changing the environment. Because as the solvent is added the electron density on carbonyl groups changes which affect the vibrational motion (vibrational frequency) and consequently the catalytic reactivity.

**Summary:**

2D IR spectroscopy method help us to unravel many dynamical processes happening in the liquid phase systems such as vibrational relaxation and frequency evolution which are the basis of understanding the structural changes inside the system. In this work we take advantage of structural sensitivity and high time resolution of this spectroscopy method to study the catalytic reactivity of a transition metal complex. Using CLS method we get an idea of how fast the environment changes. By monitoring the frequency shifts of the vibrational mode of interest, we can evaluate the reactivity of the catalyst, as the electron density is drawn away or toward the carbonyl bond. The vibrational lifetime of the carbonyl groups stretching mode is also measured using a transient grating spectroscopy coupled to 2D IR method. A shorter lifetime indicates a higher reactivity.